# Physical Origin and Generic Control of Magnonic Band Gaps of Dipole-Exchange Spin Waves in Width-Modulated-Nanostrip Waveguides

Ki-Suk Lee, Dong-Soo Han, and Sang-Koog Kim[*]

*Research Center for Spin Dynamics & Spin-Wave Devices and Nanospinics Laboratory,*

*Department of Materials Science and Engineering, Seoul National University,*

*Seoul 151-744, Republic of Korea*

We report, for the first time, on a novel planar structure of magnonic-crystal waveguides, made of a single magnetic material, in which the allowed and forbidden bands of propagating dipole-exchange spin-waves can be manipulated by the periodic modulation of different widths in thin-film nanostrips. The origin of the presence of several magnonic wide band gaps and the crucial parameters for controlling those band gaps of the order of ~10 GHz are found by micromagnetic numerical and analytical calculations. This work can offer a route to the potential application to broad-band spin-wave filters in the GHz frequency range.

The engineering of electronic band gaps in a periodic atomic structure has played a crucial role in the developments of currently advanced semiconductor technologies. Reliable manipulations of the propagations of electrons in atomic-scale periodic structures as well as electromagnetic waves (photons) in submicron- or larger-scale structures are one of longstanding fundamental issues in the field of condensed matter physics. Controlling the propagation of photons in a variety of artificially fabricated periodic structures, known as photonic crystals is a good example [1]. Owing to various applications of the photonic crystal to optical nano-devices such as photonic waveguides [2] and integrated circuits [3], the photonic crystals have been given considerable attentions. Meanwhile, in the areas of nanomagnetism and magnetization (**M**) dynamics, the magnetic counterpart of the photonic crystals, the so-called magnonic crystal (MC), is a subject of growing interest, owing to its applications to spin-wave (SW) waveguides and filters [4]. In recent years, many theoretical and experimental studies have been conducted on not only various types of MCs including one-dimensional (1D) structures such as periodic multilayers [5,6], periodic arrays of nanostrips [7], corrugated films [8,9], and comb-like [10] or serial loop structures [11], but also 2D or 3D structures [12,13,14]. In such structures, the allowed and forbidden SW modes (called magnons) are controllable by periodic structures artificially fabricated with different magnetic material parameters [5,6,15], shapes [8,9,10,11], and exchange-bias fields [16].

Despite recent advances in fundamental understandings of those MCs as well as the wave properties of excited SW modes, few studies have focused on MC waveguides composed of simple structures for its practical application to broad-band SW filters [9]. For future SW-based signal processing devices [17,18], it is necessary to find micrometer-size (or smaller) MC waveguides having simple planar structures, with controllable wide band gaps (of a few GHz) of dipole-exchange spin-waves (DESWs).

In this Letter, we report, for the first time, on a new type of simple planar-patterned thin-film nanostrip waveguides in which DESW's magnonic bands along with their wide bandgaps on the order of 10 GHz can be manipulated by periodic modulations of different widths (of a few tens of nm). The physical origin of the presence of magnonic wide band gaps in such width-modulated nanostrips and the relations of the allowed DESW modes and band gaps to geometric variation of the proposed MCs were found by micromagnetic numerical and analytical calculations.

We performed micromagnetic simulations on DESW propagations in magnetic thin-film nanostrips. We used, as a model system, 10 nm-thick Permalloy (Py) nanostrips of different widths (24 and 30 nm here for example) modulating with a periodicity $P$ ranging from 12 to 42 nm [the light-gray area in Fig. 1(a)], which were connected directly to a segment of the Py nanostrip of 10 nm thickness and 30 nm width [the yellow area in Fig. 1(a)]. The unit period of the nanostrips consists of the same Py segments with the different widths of 24 and 30 nm and with the corresponding lengths $P_1$ and $P_2$, respectively, as illustrated in Fig. 1(b). The OOMMF code (version 1.2a4) [19] was used to numerically calculate the dynamics of the **M**s of individual unit cells (size: 1.5 × 1.5 × 10 nm$^3$ [20]) interacting through exchange and dipolar forces, which code uses the Landau-Lifshitz-Gilbert equation of motion [21]. The chosen material parameters corresponding to Py are as follows: the saturation magnetization $M_s$ = 8.6 × 10$^5$ A/m, the exchange stiffness $A_{ex}$ = 1.3 × 10$^{-11}$ J/m, the damping constant $α$ = 0.01, the gyromagnetic ratio $γ$ = 2.21 × 10$^5$ m/As, and with zero magnetocrystalline anisotropy. For the local homogeneous excitation and subsequent propagation of the lowest-mode DESWs, along the length direction, with frequencies, $f_{SW}$, ranging from 0 to 100 GHz, we applied a "sine cardinal (sinc)" function, $H_y(t) = H_0 \frac{\sin[2\pi\nu_H(t-t_0)]}{2\pi\nu_H(t-t_0)}$ with $H_0$ = 1.0 T and the field frequency $\nu_H$ = 100 GHz, only to a local area of 1.5 × 30 nm$^2$ indicated by the dark-brown color shown in Fig. 1(a).

The results obtained by the fast Fourier transform (FFT) of the temporal $M_z/M_s$ evolution for DESW propagations along the $x$-axis at $y$ = 15 nm are plotted in Fig. 2 [18]. The frequency spectra clearly reveal the allowed and/or forbidden bands of the DESWs propagating through the nanostrips: the allowed bands are indicated by the colored region, and the forbidden bands, by the white region. For comparison, the fundamental mode DESWs propagating in single-width (24 and 30 nm) nanostrips are shown in Figs. 2(a) and 2(b). Obviously, there is no forbidden band except for below an intrinsic potential barrier (<14 GHz), owing to the quantization of the lowest mode of DESWs due to the geometric confinement of the nanostrip's narrow width [18,22,23,24,25]. For the different-width-modulated nanostrips, by contrast, there are several wide forbidden bands on the order of ~ 10 GHz [see Figs. 2(c), 2(d)]. Moreover, the number of forbidden bands as well as the bands' position and gap width differ according not only to $P$ but also the motif (represented by $P_1/P$). More specifically, for $P$ = 18 nm with [$P_1$, $P_2$] = [9 nm, 9 nm], two wide band gaps (11 and 16 GHz) appear in the DESW modes ranging from 14 to 100 GHz, whereas, for $P$ = 30 nm with [$P_1$, $P_2$] = [15 nm, 15 nm], five forbidden bands with smaller gap widths (3.8 ~ 8.6 GHz) exist (for more data, see Suppl. Fig. 1 [26]).

To comprehensively understand such striking band-gap features, we plotted the dispersion curves of the DESW modes in the 24 and 30 nm-wide nanostrips and in the nanostrip of [$P_1$, $P_2$] = [9 nm, 9 nm] [27] as an example. Due to the pinning of the DESWs at the longer (length direction) edges of the nanostrips, there exist certain width-modes having quantized $k_y$ values [23]. Generally, in single-width nanostrips, it is expected that several width-modes are excited [22,23,24,25] and, thus, several concave branches appear in the dispersion curves [18]. In the present simulation, however, there was a single parabolic dispersion curve, as shown in Fig. 3(a), because homogeneous DESW excitations along the width direction employed in this study led to only the lowest mode having the smallest $k_y$ value [see Fig. 1(c)]. Accordingly, one would expect that the dispersion curves for the width-modulated nanostrips be folded and have band gaps at the Brillouin zone (BZ) boundaries, similar to those found typically in a 1D periodic system [2,28]. However, the dispersion curves for [$P_1$, $P_2$] = [9 nm, 9 nm] show rather more complicated band features [Fig. 3(b)]: The band gaps occur not only at the BZ boundaries, $k_x = n\pi/P$ with integers $n$ (black dashed lines), but also at certain $k$ values, $k_x = [(2n+1)\pi \pm 1.44]/P$ (red dotted lines). The former can be explained by a periodic translation symmetry associated with the different width modulation along the DESW propagation direction, but the latter cannot be understood by such a 1D approach.

In order to quantitatively elucidate the physical origin of such different band gaps varying with different width modulation, we compared magnonic band diagrams [the thick black lines in Fig. 4 (a)] obtained numerically from micromagnetic simulations for the case of [$P_1$, $P_2$] = [9 nm, 9 nm], combined with the analytical calculation of the band structure of a single-width (27 nm) nanostrip: Note that this width is just the average of the 24 and 30 nm widths employed in the width-modulated nanostrips. For the 27 nm-wide nanostrip, the dispersion relation was analytically derived and expressed in terms of a quantized in-plane wavevector, $\kappa_m^2 = k_x^2 + k_{y,m}^2$ with integers $m$ = 1, 2, 3, etc, [18,23]. The $k_x$ and $k_y$ correspond to the longitudinal and transverse components of $\kappa_m$, respectively. The $k_{y,m}$ value can be obtained by considering the "effective" pinning [23], for example, $k_{y,m} = m \times 0.072\, nm^{-1}$ for the 27 nm-wide nanostrip [29]. In Fig. 4(a), the solid red line indicates the dispersion curves of the DESW mode with $m$ = 1, the lowest mode excited. Owing to the periodicity of the width modulation, the dispersion curves are folded at the first BZ boundary (the dashed vertical line), as shown in Fig. 4(a), and thus, these folded branches intersect with the original one at the BZ boundaries. Such a crossing of the dispersion curves indicates the 'diagonal' coupling between the two identical modes having opposite propagation vectors [30]. This diagonal coupling represents interference between the initially propagating forward mode and its backward mode reflected at the BZ boundary, resulting in the standing wave pattern of DESWs with $k_x = n\pi/P$ in the MC of $P$, and a split in the energy band (a band gap) [see the thick black lines in Fig. 4(a)] [28]. Next, Figure 4(b) shows calculations of the spatial distributions of the FFT powers of the local $M_z/M_s$ oscillations for the indicated specific frequencies selected at the top and bottom of the magnonic band for the case of $[P_1, P_2]$ = [9 nm, 9 nm]. It is evident that the origin of the first band gap [31] is the diagonal coupling between the two identical but oppositely propagating lowest-mode ($m$ = 1) DESWs, as explained above.

In addition to the first band gap at the BZ boundaries associated with the diagonal coupling, the dispersion branch of a higher-quantized width-mode ($m$ = 3), noted by the dotted orange lines in Fig. 4(a), intersects with that of the lowest mode ($m$ = 1) at $k_x = [0.5 \pm 0.2]2\pi/P$ (away from the BZ boundaries) indicated by the blue circles and the arrows in Fig. 4(a). To understand these band gaps, we consider a 2D scattering of the lowest-mode ($m$ =1) DESWs from the edge steps between the narrower- and wider-width strip segments [see Fig. 1(b)]. Such edge steps periodically arranged in the width-modulated nanostrips play a crucial role as new sources for excitations of the higher width-modes ($m$ = 3, 5, 7…) [32]. In general, those DESWs scattered from the edge steps propagate in wide angles on the $x$-$y$ plane, so that they interfere destructively with themselves. However, for the phase-matching condition of the scattered DESWs, they can interfere constructively with themselves; in other words, other higher-width-mode ($m = 3$) DESWs being propagating in the opposite direction is excited, and interacts with the initial propagating lowest mode. Consequently, the interactions between the initial lowest mode ($m = 1$) and the excited higher width-mode ($m = 3$) DESWs lead to quite complicated 2D standing wave patterns of DESWs. For $f_{SW} = 66.8$ GHz, the nodes appear in both the width- and length direction [see Figs. 4(b) and 4(c)], subsequently leading to complex 2D normal modes in thin-film nanostrips of the lateral confinements [33]. This gives rise to the anti-crossing of dispersion curves as well as band gaps [30] [see the thick black lines for the second and the third bands, and the diagonal-line-patterned blue region between them in Fig. 4(a)].

Such strong coupling between the initially propagating mode and the newly excited higher-mode, and the resulting band gaps are known as the 'mini-stopbands of electromagnetic waves in photonic crystal waveguides [2,30,34]. It is worthwhile to note that the lateral geometric confinements in the width-modulated nanostrips can also yield significant internal field inhomogeneities, as reported for a Daemon-Eschbach geometry in Refs. [24,25]. Simulation results for the dynamic **M** profiles of the normal modes between the uniform and nonuniform internal field (see Suppl. Fig. 2.) reveal that the dynamic **M** profiles in the width-modulated nanostripestrips could not be explained simply in terms of the inhomogeneity of the internal field. The complex modes of DESWs and associated band gaps in the width-modulated MCs are the result of the cooperative phenomena of the diffraction, reflection of the DESWs scattered at the edge steps of the width-modulated nanostrips and their interference with the initially propagating lowest-mode DESWs [26]. On the basis of such novel DESW's band structures, it was found that magnonic band gaps vary sensitively according to both the periodicity and the motif in width-modulated nanostrips. From an application perspective, this novel property can be implemented as an effective means of manipulating the allowed DESW modes in their propagations through such width-modulated nanostrips, a new type of SW waveguides that pass DESWs in a chosen narrow-band frequency region but filter out most DESWs having other frequencies. Moreover, these results can resolve the bottleneck of spin-wave devices – the trade-off between their speed, miniaturization, and controllability by applied magnetic fields [35].

In conclusion, we found that complex DESW's band structures and wide band gaps originate from the diagonal coupling between the identical lowest modes, as well as the coupling between the initially propagating lowest mode and the higher-quantized width-mode newly excited through the DESW's scattering at the edge steps of different width modulated nanostrips. Moreover, we found that the magnonic band-gap width, the position, and the number of band gaps are controllable by the periodicity and the motif of the different width modulation.

We express our thanks to B. Hillebrands and A. Slavin for their careful reading of this manuscript. This work was supported by Creative Research Initiatives (the Research Center for

**References:**

* corresponding author, e-mail address: sangkoog@snu.ac.kr

[1] E. Yablonovitch, Phys. Rev. Lett. **58**, 2059 (1987).

[2] J. D. Joannopoulos, R. D. Meade, and J. N. Winn, Photonic Crystals: Molding the Flow of Light, 2nd ed. (Princeton University, Princeton, NJ, 2008).

[3] N. Engheta, Science **317**, 1698 (2007).

[4] R. L. Carter *et al*., J. Appl. Phys. **53**, 2655 (1982).

[5] D. S. Deng, X. F. Jin, and R. Tao, Phys. Rev. B **66**, 104435 (2002).

[6] S. A. Nikitov, Ph. Tailhades, and C. S. Tsai, J. Magn. Magn. Mater. **236**, 320 (2001).

[7] M. P. Kostylev, A. A. Stashkevich, and N. A. Sergeeva, Phys. Rev. B **69**, 064408 (2004); M. Kostylev *et al*., Appl. Phys. Lett. **92** 132504 (2008).

[8] C. G. Sykes, J. D. Adam, and J. H. Collins, Appl. Phys. Lett. **29**, 388 (1976); P. A. Kolodin and B. Hillebrands, J. Magn. Magn. Mater. **161**, 199 (1996).

[9] A. V. Chumak *et al*., Appl. Phys. Lett. **93**, 022508 (2008).

[10] H. Al-Wahsh *et al*., Phys. Rev. B **59**, 8709 (1999).

[11] A. Mir *et al*., Phys. Rev. B **64**, 224403 (2001).

[12] J. O. Vasseur *et al*., Phys. Rev. B **54**, 1043 (1996).

[13] Yu. V. Gulyaev *et al*., JETP Lett. **77**, 567 (2003).

[14] M. Krawczyk and H. Puszkarski, Phys. Rev. B **77**, 054437 (2008).

[15] V.V. Kruglyak, and R.J. Hickena, J. Magn. Magn. Mater. **306**, 191 (2006); V.V. Kruglyak *et al.*, J. Appl. Phys. **98**, 014304 (2005).

[16] C. Bayer, M. P. Kostylev, and B. Hillebrands, Appl. Phys. Lett. **88**, 112504 (2006).

[17] R. Hertel, W. Wulfhekel, and J. Kirschner, Phys. Rev. Lett. **93**, 257202 (2004); T. Schneider *et al.*, Appl. Phys. Lett. **92**, 022505 (2008); K.-S. Lee and S.-K. Kim, J. Appl. Phys. **104**, 053909 (2008).

[18] S. Choi, K.-S. Lee, K. Y. Guslienko and S.-K. Kim, Phys. Rev. Lett. **98**, 087205 (2007).

[19] A version of the OOMMF code used is 1.2a4. See http://math.nist.gov/oommf.

[20] Simulation results using $1.5 \times 1.5 \times 2.5$ nm$^3$ are in good agreements with those using $1.5 \times 1.5 \times 10$ nm$^3$.

[21] L. D. Landau and E. M. Lifshitz, Phys. Z. Sowjet. **8**, 153 (1935); T. L. Gilbert, Phys. Rev. **100**, 1243A (1955).

[22] T. W. O'Keeffe and R. W. Patterson, J. Appl. Phys. **49**, 4886 (1978); R. Arias and D. L. Mills Phys. Rev. B **70**, 094414 (2004); V. E. Demidov and S. O. Demokritov, Phys. Rev. **B 77**, 064406 (2008).

[23] K. Yu. Guslienko *et al.*, Phys. Rev. B **66**, 132402 (2002); K. Yu. Guslienko and A. N. Slavin, Phys. Rev. B **72**, 014463 (2005).

[24] M. P. Kostylev *et al.*, Phys. Rev. B **76**, 054422 (2007).

[25] V. E. Demidov et al. Appl. Phys. Lett. **92**, 232503 (2008).

[26] See EPAPS Document No. XXX for two figure files and five movie files. For more information on EPAPS, see http://www.aip.org/pubservs/epaps.html.

[27] Dispersion curves were obtained from the 2D FFTs of the temporal $M_z/M_s$ oscillations along the x-axis (x = 501 ~ 1500 nm) at y = 15 nm (the dashed line on the nanostrip MC in Fig. 1).

[28] C. Kittel, Introduction to Solid State Physics, 7th ed. (Wiley, New York, 1996).

[29] In the calculation of the $k_{y,m}$ values for the 27 nm-width nanostrip having a homogeneous internal field distribution across the strip width, the effective "pinning" parameter expressed by Eq. (5) in Ref. [23] was used. The exact values of $k_{y,m}$ can be obtained using Eq. (6) in Ref [24]. However, the approximate analytical calculation described in Ref. [23] is in a good agreement with the result obtained with Eq. (6) in Ref. [24] for the single-width nanostrip of longitudinally saturated magnetizations.

[30] J.-M. Lourtioz *et al.*, Photonic Crystals: Towards Nanoscale Photonic Devices (Springer, Berlin, 2005).

[31] The top of the first allowed band ($f_{SW}$ = 25.6 GHz) and the bottom of the second allowed band ($f_{SW}$ = 36.4 GHz) appear at the first BZ boundary, $k_x = \pi/P$.

[32] Since the edge steps are symmetric with respect to the x-axis (mirror plane), higher-quantized width-modes have a mirror symmetry in the profiles along the width axis (y-axis); in other words, they have odd-number m values (i.e., m = 3, 5, 7…).

[33] M. Grimsditch *et al.*, Phys. Rev. B **69**, 174428 (2004); M. Grimsditch *et al.*, Phys. Rev. B **70**, 054409 (2004).

[34] S. Olivier *et al.*, Phys. Rev. B **63**, 113311 (2001).

[35] S. V. Vasiliev *et al.*, J. Appl. Phys. **101**, 113919 (2007).

**Figure captions**

FIG. 1. (Color online) (a) Geometry and dimensions of proposed nanostrip MCs with periodic modulation of different strip widths. The initial **M**s point in the –*x* direction, as indicated by the black arrow. The dark-brown and yellow areas indicate the SW generation and waveguide component, respectively. (b) The unit period of $P = P_1 + P_2$, where $P_1$ and $P_2$ are the segment lengths of 24 and 30 nm widths, respectively. (c) Temporal evolution of spatial $M_z/M_s$ distribution excited by "sinc" function field with $H_0 = 1.0$ T applied along *y*-axis to only dark-brown area.

FIG. 2. (Color online) Frequency spectra obtained from FFTs of $M_z/M_s$ oscillation along *x*-axis at *y* = 15 nm, for single-width nanostrips (24 and 30 nm) and for MCs of different [$P_1$, $P_2$] values noted. The vertical dashed orange lines indicate the boundary between the single-width nanostrip waveguide [the yellow area in Fig. 1(a)] and the MC of the width-modulated nanostrip [the light-gray area in Fig. 1(a)].

FIG. 3. (Color online) (a) Dispersion curves for DESWs propagating through single-width nanostrips of 24 and 30 nm. (b) Dispersion curves of DESWs existing in the MC of [$P_1$, $P_2$] = [9 nm, 9 nm] within nanostrip area only, from *x* = 500 to 1500 nm, obtained from FFTs of temporal $M_z/M_s$ oscillations along the *x*-axis at *y* = 15 nm. The black dashed lines indicate the Brillouin zone boundaries $k = n\pi/P$, where $n = 0, \pm1, \pm2, \ldots$, and the red dotted lines denote certain $k$ values, $k_x = (2n+1)\pi/P \pm 0.08$ at which the forbidden band gaps occur.

FIG. 4. (Color online) (a) Comparison of magnonic band structure (black thick curves) of nanostrip-type MC of $[P_1, P_2]$ = [9 nm, 9 nm] obtained from micromagnetic simulations and that of a single-width nanostrip of 27 nm width, for two different modes of $m$ = 1 (solid red) and $m$ = 3 (dotted orange) obtained from the analytical form of Eq. (1) in Ref. [18]. (b) Perspective view of the FFT power distributions of local $M_z/M_s$ oscillations for specific frequencies for the top and bottom of allowed bands, as indicated by orange horizontal lines in the dispersion curves shown in Fig. 3(b). (c) Cross-sectional FFT power profiles of standing wave modes in width direction ($y$-direction). The red (green) line indicates the standing wave profile in the width direction at the center of the 30 nm- (24 nm-) wide segment.

**FIG.1**

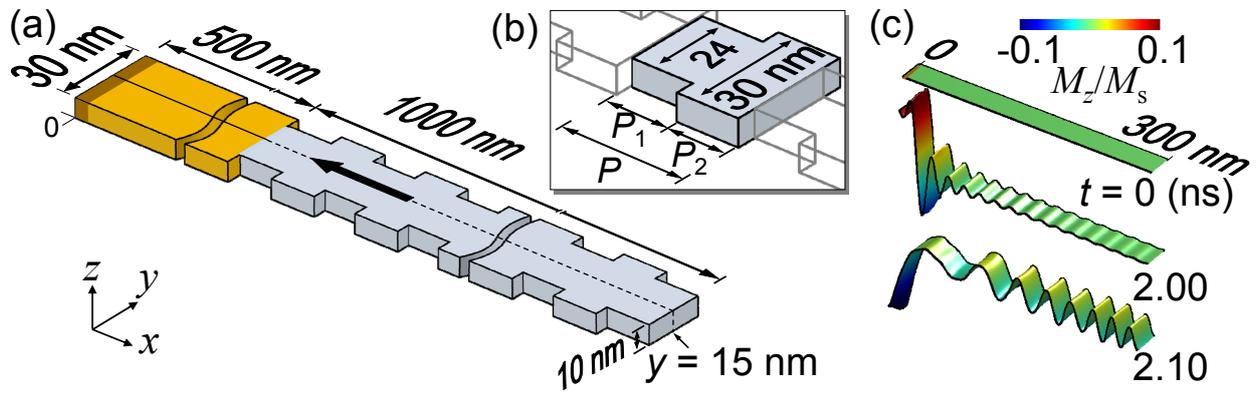

**FIG.2**

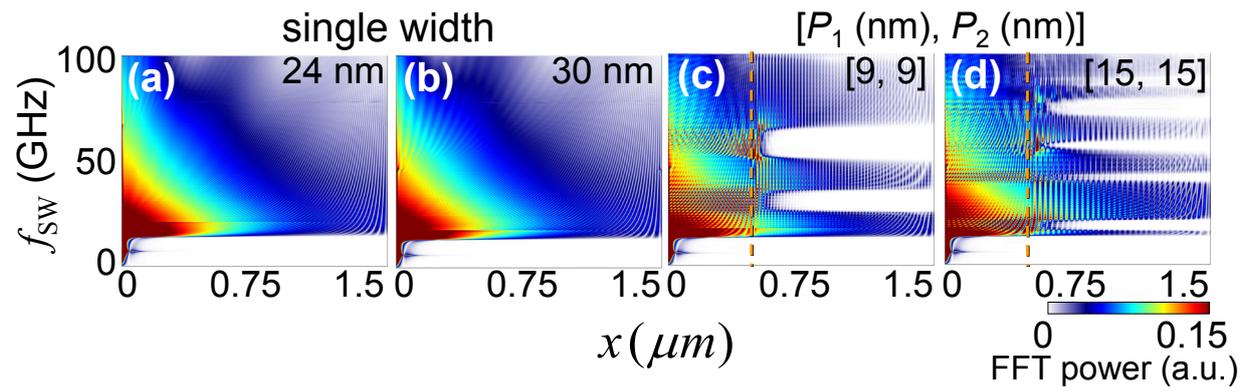

FIG.3

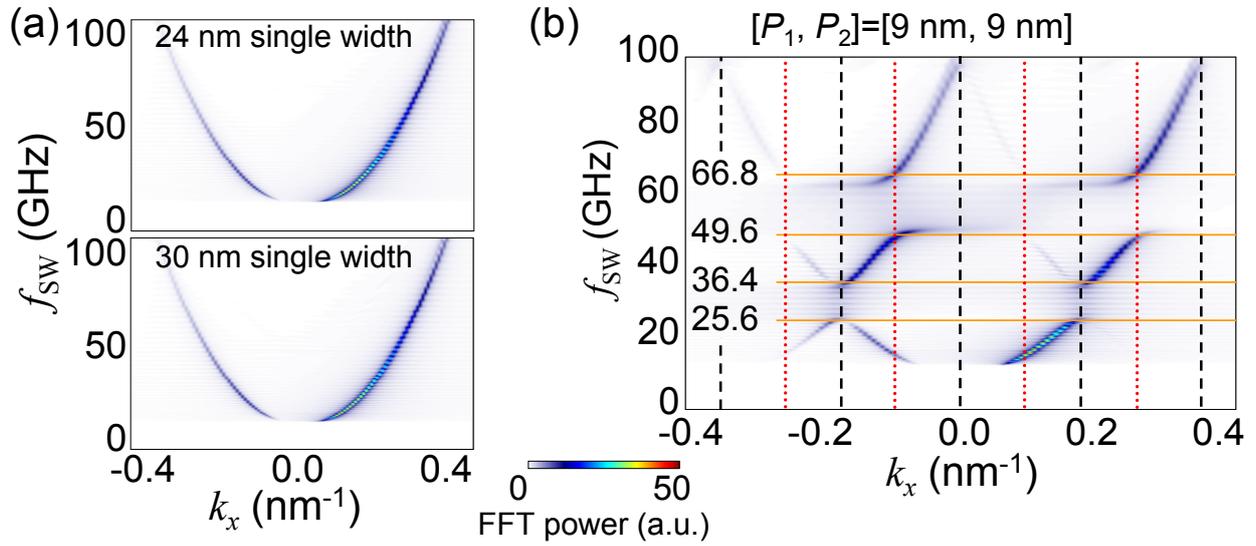

FIG.4

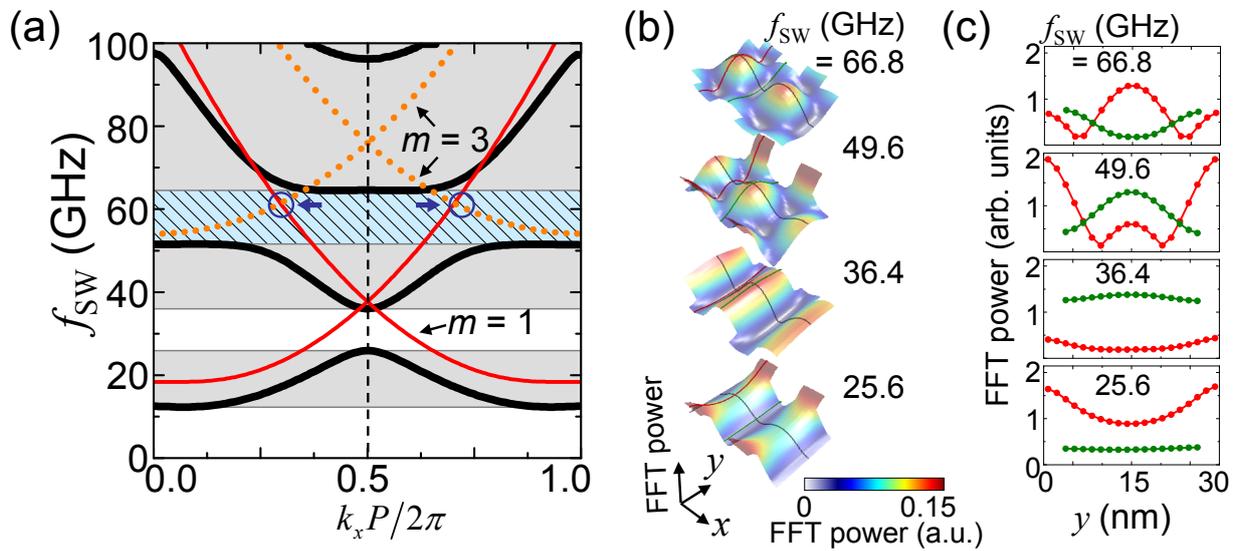

# Supplementary Document

**The non-linearity of DESWs in the simulations**

We compared the magnonic band structures for various intensities of the initially propagating DESWs, as obtained from micromagnetic simulations with various $H_0$ values ranging from 10 Oe to 1.0 T. The results reveal that the noticeable magnonic band structure is not affected by the excitation field strength up to 1.0 T. In other words, the non-linearity of DESWs is ignorable in the present simulations. Consequently, a nonlinear effect of DESWs is not the key origin of the band gaps found at certain $k$ values, away from the BZ boundaries.

**SUPPL. FIG. 1**

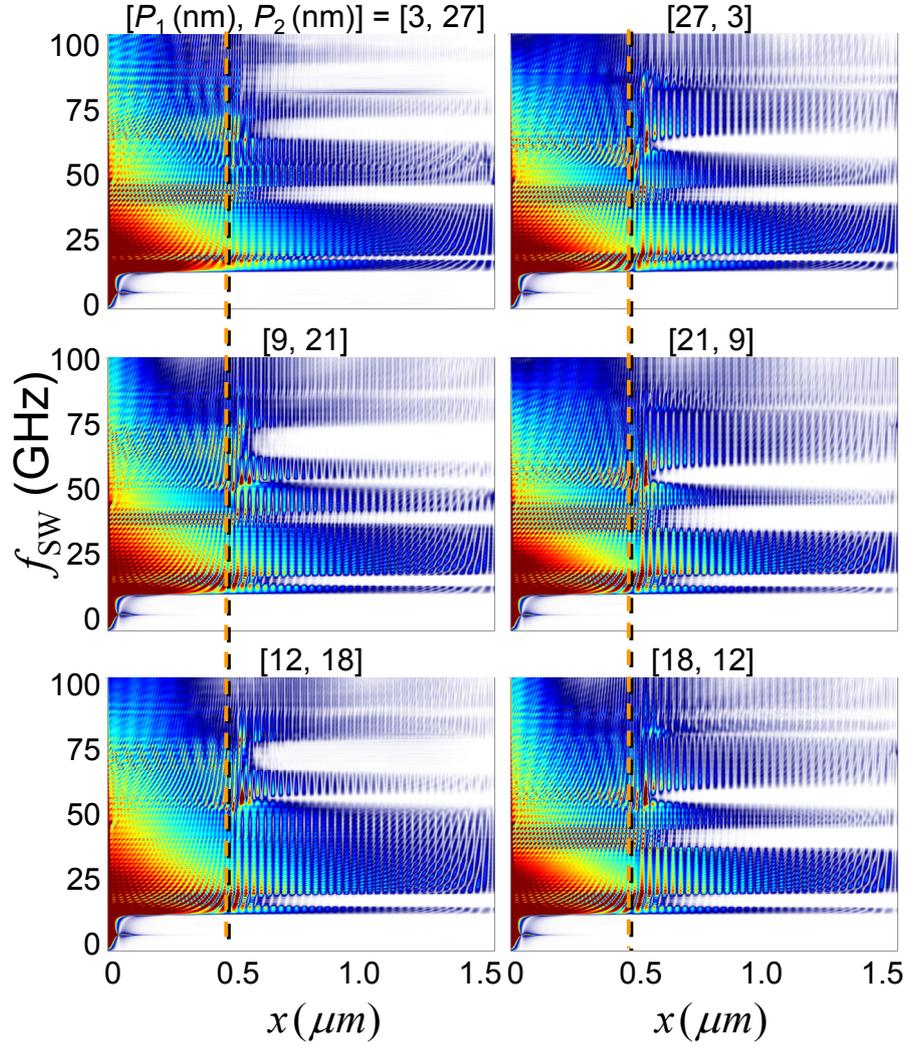

**SUPPL. FIG. 1**. Frequency spectra obtained from fast Fourier transform (FFT) of $M_z/M_s$ oscillation along the *x*-axis at $y = 15$ nm, for cases of the same $P = 30$ nm but different $P_1/P$ values indicated. The vertical dashed orange lines indicate the boundary between the single-width nanostripe waveguide [the yellow area in Fig. 1(a)] and the MC waveguide of a width-modulated nanostrip [the light-gray area in Fig. 1(b)].

**SUPPL. FIG. 2**

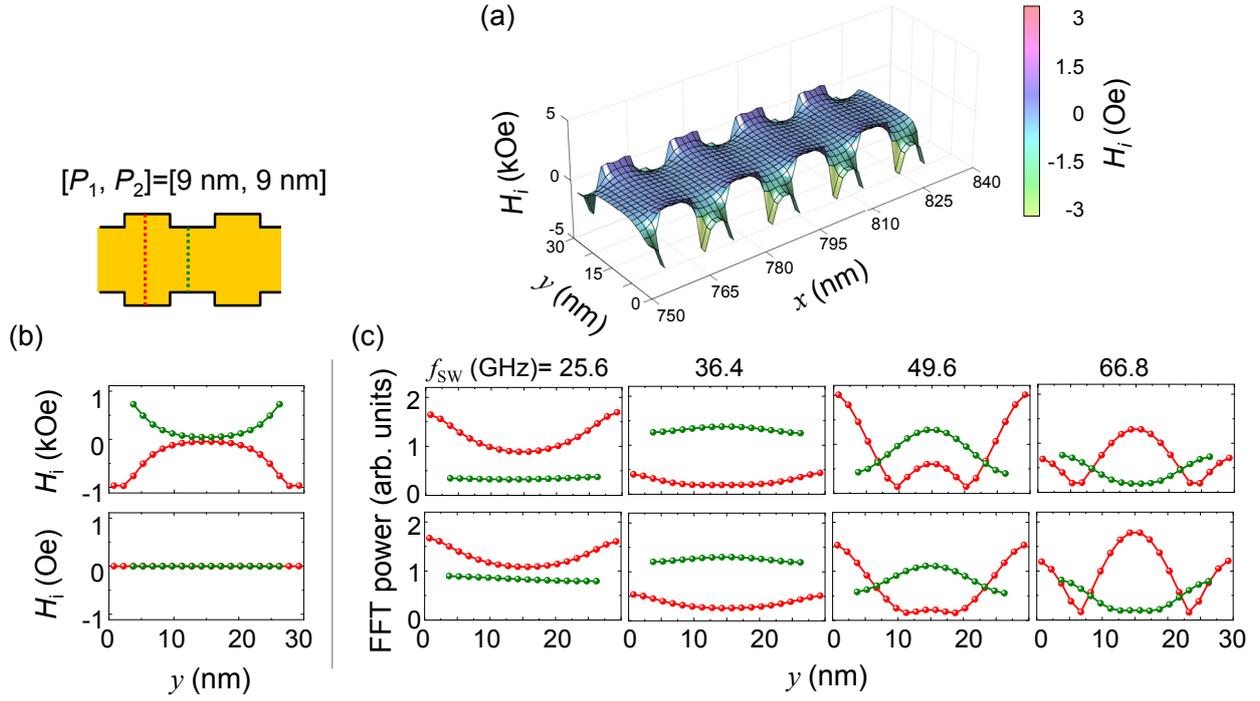

**SUPPL. FIG. 2**. (a) Ground-state internal field ($H_i$) distribution in the MC $[P_1, P_2]$ = [9 nm, 9 nm] under no applied field. (b) Upper panel: Inhomogenous internal field profiles along the width direction, for the 30-(red) and 24-nm (green) wide strip segments, as indicated by the red and green lines in the inset of (b). Lower panel: Homogenous zero internal field along the width direction for the 30-(red) and 24-nm (green) wide strip segments after applying a specially designed external filed distribution. (c) Cross-sectional profiles of the FFT power of the excited SW modes along the width direction at the center of the 30 nm- and 24 nm-wide strip segments, obtained from micromagnetic simulations for both cases of the inhomogeneous (the first row) and homogeneous (the second row) $H_i$ distributions.

# Supplementary Movies

**Supplementary Movie 1.** Animation of the temporal evolution of the spatial $M_z/M_s$ distribution excited by a "sinc" function field with $H_0 = 1.0$ T applied along the *y*-axis to only dark-brown area in Fig. 1(a). The color and height of the surfaces indicate the local out-of-plane **M** components, according to the same scale noted in Fig. 1(c).

**Supplementary Movies 2a, 2b, 2c, and 2d** represent animations of the temporal evolution of the propagating DESW modes in nanostrip-type MC with $[P_1, P_2] = [9$ nm, $9$ nm$]$ for given frequencies of 20, 30, 40, and 55 GHz, respectively. The colors indicate the local out-of-plane **M** component normalized by the saturation value, i.e., $M_z/M_s$ according to the same scale noted in Fig. 1(c).